\documentclass[11pt]{article}
\usepackage{amsfonts,epsfig}
\usepackage{amsfonts}
\usepackage{amsmath}
\usepackage{amssymb}
\usepackage{graphicx}%
\begin{document} 
\noindent

\vspace*{0cm}

\begin{center}
{\setlength{\baselineskip}{1.0cm}{\ {\Large \textbf{AN EXACTLY-SOLVABLE THREE-DIMENSIONAL \\[1ex]
NONLINEAR QUANTUM OSCILLATOR}} \textbf{ }} \vspace*{1cm} }

{\large \textsc{A. Schulze-Halberg}$^{\dagger}$ and \textsc{J.R.
Morris}$^{\ddagger}$}
\end{center}

\noindent\newline

$\dagger$ Department of Mathematics and Actuarial Science, Indiana University
Northwest, 3400 Broadway, Gary IN 46408, USA, e-mail: axgeschu@iun.edu, xbataxel@gmail.com

\bigskip

$\ddagger$ Department of Physics, Indiana University Northwest, 3400 Broadway,
Gary IN 46408, USA, e-mail: jmorris@iun.edu \newline

\begin{abstract}

Exact analytical, closed-form solutions, expressed in terms of special
functions, are presented for the case of a three-dimensional nonlinear quantum
oscillator with a position dependent mass. This system is the generalization
of the corresponding one-dimensional system, which has been the focus of
recent attention. In contrast to other approaches, we are able to obtain
solutions in terms of special functions, without a reliance upon a
Rodrigues-type of formula. The wave functions of the quantum oscillator have
the familiar spherical harmonic solutions for the angular part. For the
s-states of the system, the radial equation accepts solutions that have been
recently found for the one-dimensional nonlinear quantum oscillator, given in
terms of associated Legendre functions, along with a constant shift in the
energy eigenvalues. Radial solutions are obtained for all angular momentum
states, along with the complete energy spectrum of the bound states. 

\end{abstract}


\section{Introduction}

An interesting one-dimensional model of a nonlinear quantum oscillator has
been studied recently by Cari\~{n}ena, Ra\~{n}ada and Santander \cite{cari}.
This model presents two interesting features, namely nonlinearity of the
potential $V(x)$ and the presence of a position dependent mass $M(x)$,
producing a position dependent spring \textquotedblleft
constant\textquotedblright\ $K(x)\propto M(x)$ in the oscillator potential.
This feature allows the existence of solutions for the classical nonlinear
oscillator that have a simple harmonic form. The quantum mechanical version of
the model admits exact solutions for the wave functions and the energy
eigenvalues, which were obtained by the authors of Ref. \cite{cari}. This was
achieved through the use of a factorization method, and solutions for the wave
functions could be represented by means of a Rodrigues-type of formula. The
intriguing features and the solvabilty of this model makes it an attractive
instructional example that could yield insights into other physical systems
and mathematical methods. Consequently, this same one-dimensional nonlinear
quantum oscillator model was further studied by Schulze-Halberg and
Morris \cite{axel1}, where the exact solutions are expressed in closed-form in
terms of special functions, without the need to use the factorization method
or Rodrigues-type of formula of Ref. \cite{cari}. The full solution spectrum is
obtained, and is seen to exhibit orthogonality and normalizability, with a
proper reduction to the harmonic oscillator model in the appropriate limit.

\bigskip

\ \ Here we focus attention on the three-dimensional generalization of the 1D
model of Refs. \cite{cari},\cite{axel1}. We note that this 3D version of the
model takes into account not only vibrational degrees of freedom, but
rotational ones, as well. That is, we consider all possible angular momentum
states, not only $s$ states which have a vanishing centrifugal term in the
effective potential for the radial motion. The resulting energy spectrum
therefore involves both a vibrational quantum number $n$ and an angular
momentum quantum number $L$. The effective potential for this 3D system is of
a more complicated form, but has spherical symmetry, and presents a more
realistic representation of a system with variable mass $M(r)$ that can both
vibrate and rotate.

\bigskip

\ \ Position dependent masses can arise in condensed matter settings, as
mentioned in Ref. \cite{cari}, and in theories of gravitation, such as
scalar-tensor theories (see, for example, Ref. \cite{F-M}), where a fixed,
constant mass in one conformal frame becomes a position dependent mass in a
different conformal frame, as originally pointed out by Dicke \cite{Dicke62}.
An example of this, analyzed by Dicke \cite{Dicke62}, is seen in the
prototypical Brans-Dicke theory \cite{Brans-Dicke}, where a test particle with
constant mass $m_{0}$ in the \textquotedblleft Jordan frame\textquotedblright%
\ representation of the theory has a position dependent mass $m[(\phi(x^{\mu
})]=A[\phi(x^{\mu})]m_{0}$ in the \textquotedblleft Einstein
frame\textquotedblright\ representation. Here, the function $A(\phi)$ of the
scalar field $\phi(x^{\mu})$ connects the metrics in the two different
conformal frames: $\tilde{g}_{\mu\nu}(x^{\alpha})=A^{2}(\phi)g_{\mu\nu
}(x^{\alpha})$, with $\tilde{g}_{\mu\nu}$ representing the Jordan frame metric
and $g_{\mu\nu}$ the Einstein frame metric. Since the scalar field $\phi$ is,
in general, a function of spacetime coordinates, the mass of the test particle
in the Einstein frame becomes spacetime position dependent. Therefore, well
motivated reasons exist  for considering theories with position dependent mass.

\bigskip

\ \ Here we consider the nonlinear quantum oscillator studied in
Refs. \cite{cari} and \cite{axel1}, generalized to three dimensions. We follow
the procedure of Ref. \cite{cari} of formulating the classical Hamiltonian in
such a way as to allow us to pass to the quantum Hamiltonian, with a
quantization procedure resembling canonical quantization. We arrive at the 3D
time independent Schr\"{o}dinger equation, which we separate into angular and
radial parts. The angular solutions are the usual spherical harmonics. We then
focus on the more difficult problem of solving the radial equation. This is
achieved by presenting solutions in the form of special functions. The entire
bound state energy spectrum is obtained in terms of vibrational and rotational
quantum numbers.

\section{The classical system and its quantization}
In this section we derive a quantum model featuring a nonlinear oscillator potential in one and three dimensions. 
To this end, we start out from the classical Lagrangian and construct the quantum-mechanical Hamiltonian. In 
addition, we will briefly comment on solutions of the classical system.

\subsection{The one-dimensional case}
\ \ Our starting point of the analysis carried out in this note is the following, one-dimensional oscillator potential that 
was first studied in \cite{cari}:
\begin{eqnarray}
V&=&\frac{1}{2}%
~m~\alpha^2~\frac{x^{2}}{(\lambda~x^{2}+1)}, \label{pot1}%
\end{eqnarray}
where $x$ is a real, scalar variable and $m,\alpha>0$ stand for constants, $m$ representing a mass
parameter. The classical Lagrangian $L$ for the system associated with the potential (\ref{pot1}) is given by%
\begin{eqnarray}
L&=&\frac{1}{2}\left(  \frac{1}{\lambda~x^{2}+1}\right)  \left(  m~\dot{x}%
^{2}-m~\alpha^{2}x^{2}\right)  \nonumber \\[1ex]
&=&\frac{1}{2}\left(  \frac{1}{\lambda~x^2+1%
}\right)  \left(  m~\dot{x}^{2}-g~x^{2}\right),  \label{1}%
\end{eqnarray}
note that the dot stands for differentiation with respect to time and that the abbreviation $g=m\alpha^2$ was introduced. The Lagrangian (\ref{1}) is seen to have a noncanonical kinetic term, and we
identify a position-dependent mass $M=M(x)$, related to the mass parameter $m$ by%
\begin{eqnarray}
M&=&\frac{m}{(\lambda~x^2+1)}~~=~~\frac{g}{\alpha^2~(\lambda~x^2+1%
)}. \label{3}
\end{eqnarray}
Let us note that the equation of motion following from our Lagrangian (\ref{1}) is%
\begin{eqnarray}
(\lambda~x^{2}+1)~\ddot{x}-(\lambda~ x)~\dot{x}^{2}+\alpha^{2}~x&=&0. \label{4}%
\end{eqnarray}
Now, in order to obtain the classical Hamiltonian for the system governed by the Lagrangian (\ref{1}), we extract 
its kinetic energy $T$:
\begin{eqnarray}
T&=&\frac{m~\dot{x}^{2}}{2~(\lambda~x^2+1)}~~=~~\frac{1}{2}~
M~\dot{x}^{2}~~=~~\frac{p^{2}}{2~M},\label{5}%
\end{eqnarray}
where $p$ denotes the canonical momentum, given by $p=M \dot{x}$. This leads us to define a modified 
momentum variable $P$, determined by the condition
\begin{eqnarray}
\frac{P}{\sqrt{m}}&=&\frac{p}{\sqrt{M}}. \label{mcond}
\end{eqnarray}
This relation renders the kinetic energy (\ref{5}) in the form
\begin{eqnarray}
T&=&\frac{P^{2}}{2~m}~~=~~\frac{p^{2}}{2~M}, \label{7}%
\end{eqnarray}
such that the classical Hamiltonian for our system can now be written down explicitly as the sum of 
kinetic energy (\ref{7}) and potential
\begin{eqnarray}
H&=&\frac{1}{2~m}~P^{2}+V \label{ham} \\[1ex]
&=&\frac{1}{2~m}\left(  \frac{m}{M}\right)
p^{2}+V \nonumber \\[1ex]
&=&\frac{1}{2~m}~(\lambda~x^2+1)~p^{2}+V.\label{8}%
\end{eqnarray}
Note that the potential $V$ is defined in (\ref{pot1}).

\paragraph{Quantization:}\ \ The transition to the quantum-mechanical context is done by introducing the 
momentum operator $p =-i~ \hbar~ d/dx$, which converts (\ref{ham}) into the Hamiltonian operator $\hat{H}$ 
of our quantum system:
\begin{eqnarray}
\hat{H} &=& \frac{1}{2~m}~P~P+V \nonumber \\[1ex]
&=& \frac{1}{2~m}~\sqrt{\lambda~x^2+1}~p~\sqrt{\lambda~x^2+1}~p+V \nonumber \\[1ex]
&=&-\frac{\hbar^{2}}{2~m}~
\sqrt{\lambda~x^2+1}~ \frac{d}{dx} ~\sqrt{\lambda~x^2+1} ~\frac{d}{dx}+V  \nonumber\\[1ex]
& =&-\frac{\hbar^{2}}{2~m}\sqrt{\lambda~x^{2}+1}~\left[  \left(  \frac{\lambda
x}{\sqrt{\lambda~x^{2}+1}}~\frac{d}{dx}+\sqrt{\lambda~x^{2}+1} ~\frac{d^2}{dx^2}
\right)  \right]+V  \nonumber \\[1ex]
&=& -\frac{\hbar^{2}}{2~m}\left[  (\lambda~x^{2}+1)~\frac{d^2}{dx^2}+\lambda~x~\frac{d}{dx}\right] + 
\frac{1}{2}~m~\alpha^2~\frac{x^{2}}{(\lambda~x^{2}+1)}, \label{qham}
\end{eqnarray}
note that in the last step we have used the explicit representation of our potential (\ref{pot1}). The Hamiltonian presented 
in Ref. \cite{cari} (see above Eq(6) of that paper) coincides with our finding (\ref{qham}).

\subsection{The three-dimensional case}

\ \ We can extend the previous to three dimensions in a straightforward way, replacing the scalar variable $x$ by 
a coordinate triple, denoted by $\vec{r}$. We will work in spherical coordinates $(r,\theta,\varphi)$, defined in the 
usual way, such that the classical Lagrangian (\ref{1}) is converted its three-dimensional counterpart as follows:
\begin{eqnarray}
L&=& \frac{1}{2}\left(  \frac{1}{\lambda~r^2+1 }\right)  \left(  m \left\Vert\vec{v}\right\Vert^{2}
-g~r^2\right), \label{l3d}
\end{eqnarray}
where the abbreviation $\vec{v} = \dot{\vec{r}}$ was used and $\Vert \cdot \Vert$ stands for the euclidean norm in 
$\mathbb{R}^3$, note further that $\Vert \vec{r} \Vert = r$. The position-dependent mass constructed in (\ref{3}) 
maintains its form, except for the change of variable:
\begin{eqnarray}
M&=&\frac{m}{(\lambda~r^2+1)}~~=~~\frac{g}{\alpha^2~(\lambda~r^2+1%
)}. \nonumber
\end{eqnarray}
The equation of motion generated by the Lagrangian (\ref{l3d}) is very similar to its one-dimensional 
counterpart (\ref{4}):
\begin{eqnarray}
(\lambda~r^2+1)~\overset{\cdot\cdot}{\vec{r}}-(\lambda~\vec{r})~\Vert\vec{v}\Vert^2+\alpha^{2}~\vec{r}&=&0\label{12}%
\end{eqnarray}
Following the procedure shown for the one-dimensional context, we incorporate the mass condition (\ref{mcond}) and 
arrive at the following three-dimensional classical Hamiltonian
\begin{eqnarray}
H&=&\frac{1}{2~m}~\Vert\vec{P}\Vert^{2}+V \nonumber \\[1ex]
&=&\frac{1}{2~m}~(\lambda~r^2+1)~\Vert \vec{p} \Vert^2+V \nonumber \\
&=& \frac{1}{2~m}~(\lambda~r^2+1)~\Vert \vec{p} \Vert^2+\frac{1}{2}~m~\alpha^2~\frac{r^{2}}{(\lambda~r^2+1)}, \label{14}
\end{eqnarray}
where $V$ is taken from the explicit form (\ref{pot1}) with $x^2$ replaced by $r^2$. As a consequence, the potential 
is now spherically symmetric.

\paragraph{Quantization:}\ \ As before, the quantum-mechanical Hamiltonian operator $\hat{H}$ is obtained by 
expressing the 
classical momentum $\vec{P}$ through the operator $p=-i~\hbar~\nabla$. This renders the Hamiltonian operator 
$\hat{H}$ in the following form:
\begin{eqnarray}
\hat{H} &=&  \frac{1}{2~m}~\vec{P}~\vec{P}+V \nonumber \\[1ex]
&=& \frac{1}{2~m}~\sqrt{\lambda~r^2+1}~p~\sqrt{\lambda~r^2+1}~p+V \nonumber \\[1ex]
&=&-\frac{\hbar^{2}}{2~m}~
\sqrt{\lambda~r^2+1}~ \nabla ~\sqrt{\lambda~r^2+1} ~\nabla+V  \nonumber\\[1ex]
& =&-\frac{\hbar^{2}}{2~m}\sqrt{\lambda~r^2+1}~ \left(  \frac{\lambda
~r}{\sqrt{\lambda~r^2+1}}~\frac{\partial}{\partial r}+\sqrt{\lambda~r^2+1} ~\nabla^2
\right)  +V  \nonumber \\[1ex]
&=& -\frac{\hbar^{2}}{2~m}\left[  (\lambda~r^2+1)~\nabla^{2}+\lambda
~r~\frac{\partial}{\partial r}\right]  +\frac{1}{2}~m~\alpha^2~\frac{r^{2}}{(\lambda~r^2+1)}, \label{h3d}
\end{eqnarray}
where the explicit form of $V$ was taken from (\ref{14}).

\subsection{Classical Solutions}
Before we start analyzing the three-dimensional quantum model, let us briefly comment on the solutions of the 
classical equations of motion.

\paragraph{The one-dimensional case.} Our equation of motion (\ref{4}) admits the following solution 
\cite{Mathews} \cite{cari}:
\begin{eqnarray}
x &=& A~\sin(\omega~t+\phi), \label{osc}
\end{eqnarray}
where $A$, $\omega$ and the phase $\phi$ are real-valued constants, related through the constraint
\begin{eqnarray}
A^{2}&=&\frac{1}{\lambda}\left(  \frac{\alpha^{2}}{\omega^{2}}-1\right)  \label{16}
\end{eqnarray}
The solution (\ref{osc}) has the form of a conventional harmonic oscillator. Note that we can put the potential into 
harmonic oscillator shape: from (\ref{pot1}), (\ref{3}), and (\ref{16}) we can rewrite the potential as
\begin{eqnarray}
V&=&\frac{1}{2}~M~\alpha^{2}~x^{2} \nonumber \\[1ex]
&=& \frac{1}{2}~M \left( 1+ \lambda~A^{2}\right) ~\omega^{2}~x^{2} \nonumber \\[1ex]
&=& \frac{1}{2}~K~x^{2}, \label{pot3}%
\end{eqnarray}
where we have employed the abbreviation
\begin{eqnarray}
K&=&M~\left(  1+\lambda~A^{2}\right) ~ \omega^{2} \nonumber \\[1ex]
&=& m~\omega^{2}~\frac
{(1+\lambda~A^{2})}{(\lambda~x^2+1)}. \label{K}
\end{eqnarray}
Note that in the last step we used relation (\ref{3}), where $x$ is replaced by $r$. Expression $K$ could be seen as 
an effective spring \textquotedblleft constant\textquotedblright.

\paragraph{The three-dimensional case.}\ \ The equation of motion (\ref{l3d}) admits a solution that resembles its 
one-dimensional counterpart (\ref{osc}):
\begin{eqnarray}
\vec{r}&=&\vec{A}~\sin(\omega ~t+\phi), \label{17}
\end{eqnarray}
where $\vec{A}$ is a constant amplitude, $\omega$ denotes the oscillation frequency and $\phi$ is a phase. As in 
the previous case, the latter parameters are subject to the constraint of (\ref{16}),
with $A=\Vert\vec{A}\Vert$. This is not surprising, since this solution represents a
one-dimensional oscillation with respect to some fixed axis in the $(x,y,z)$ space. One can
rotate the coordinate system so that the motion takes place along the $x$
axis, with $y=z=0$, in which case the three-dimensional solution degenerates to the single-variable 
situation. The function (\ref{17}) simply represents an $s$-state where the
oscillator undergoes linear motion with no rotation.

\bigskip

\ \ If rotational motion is taken into account, the situation is more
complicated. Let us assume motion with vibration and rotation within the $x$-$y$
plane, and use polar coordinates $(r,\theta)$ to describe the motion. We make use of the relations arising from 
twofold differentiation of $\vec{r}$ with respect to time $t$:
\begin{eqnarray}
\vec{r}~=~r~\hat{r} 
\qquad \qquad
\vec{v}~=~\dot{r}~\hat{r}+r~\dot{\theta}~\hat{\theta}
\qquad \qquad
 \vec{a}~=~(\ddot{r}-r~\dot{\theta}^{2})\hat{r}+(r~\ddot{\theta
}+2~\dot{r}~\dot{\theta})~\hat{\theta}, \label{polar}%
\end{eqnarray}
where $\hat{r}$ and $\hat{\theta}$ stand for unit vectors in polar coordinates, that is,
\begin{eqnarray}
\hat{r} ~=~ \frac{\partial \vec{r}}{\partial r}~\left\Vert \frac{\partial \vec{r}}{\partial r} \right\Vert^{-1}
\qquad \qquad
\hat{\theta} ~=~ \frac{\partial \vec{r}}{\partial \theta}~\left\Vert \frac{\partial \vec{r}}{\partial \theta} \right\Vert^{-1}
\end{eqnarray}
Our equation of motion (\ref{l3d}) now yields that $r \ddot{\theta
}+2 \dot{r} \dot{\theta}=0$, which implies $r^{2}\dot{\theta
}=C=$ constant. Using this, the radial part of equation (\ref{l3d}) can be written as%
\begin{eqnarray}
(\lambda~r^2+1)\left(  \ddot{r}-\frac{C^{2}}{r^{3}}\right)  -\lambda
~r\left(  \dot{r}^{2}+\frac{C^{2}}{r^{2}}\right)  +\alpha^{2}~r&=&0\label{radial}%
\end{eqnarray}

This coincides with the one-dimensional equation of motion (\ref{4}) for zero angular momentum $C=0$, that is, 
$\dot{\theta}=0$. However, for $C\neq0$ the motion is more complicated.

\section{The three-dimensional quantum system}
In the single-variable case, the quantum Hamiltonian (\ref{qham}) was shown to admit a discrete spectrum and 
an orthogonal set of corresponding eigenfunctions, both of which were obtained in closed form \cite{cari,axel1}. 
In this section we will show that the same is possible in the three-dimensional situation, where the dynamics 
is governed by the Hamiltonian (\ref{h3d}).

\subsection{Separation of variables}
We start out by setting up the stationary Schr\"odinger equation that is generated by our three-dimensional 
Hamiltonian (\ref{h3d}). To this end, let us recall that the corresponding potential is extracted from 
(\ref{pot1}), where $x^2$ is replaced by $r^2$:
\begin{eqnarray}
V&=&\frac{1}{2}~g~\frac{r^{2}}{(\lambda~r^2+1)}, \label{pot3d}
\end{eqnarray}
where $g=m~\alpha^2$. This potential enters in our Schr\"odinger equation, which reads
\begin{eqnarray}
\left(\hat{H}-E \right) \Psi &=& 0 \nonumber \\
\left[  (\lambda~r^2+1)~\nabla^{2}+\lambda ~r~\frac{\partial}{\partial r}+\frac{2~m}{\hbar^{2}}~\left(  E-V\right)
\right]  \Psi&=&0.\ \ \ \ \ \label{20}%
\end{eqnarray}
Note that the solution $\Psi$ is allowed to depend on all three spatial variables. Now, the usual process of 
variable separation in spherical coordinates \cite{messiah} renders $\Psi$ as a product of a radial and an angular 
solution, that is, $\Psi(r,\varphi,\theta) = R(r)~Y(\varphi,\theta)$. Taking the separation constant as 
$L(L+1)$ for a nonnegative integer $L$, the two functions $R$ and $Y$ are found to satisfy the following equations:
\begin{eqnarray}%
(\lambda~r^2+1)~R''+\left(\frac{2}{r}+3~\lambda~r \right) R'+
\left[\frac{2~m}{\hbar^2} ~\left(E-V \right)-\frac{(\lambda~r^2+1)~L~(L+1)}{r^2} \right] R&=& 0 \label{radgl} \\[1ex]
\left\{  \dfrac{1}{\sin(\theta)}~\frac{\partial}{\partial \theta} \left[\sin(\theta)~\frac{\partial}{\partial \theta}\right]+
\dfrac{1}{\sin^{2}(\theta)}~\frac{\partial^2}{\partial \varphi^{2}}+L~(L+1)\right\}  Y&=& 0, \nonumber \\
\label{28}%
\end{eqnarray}
where the prime denotes differentiation with respect to the radial variable $r$. The two-dimensional angular equation 
(\ref{28}) can be solved in terms of spherical harmonics \cite{messiah}, 
usually denoted by $Y=Y_{l}^{\mu}(\theta,\varphi)$, where the integer $\mu$ satisfies $\mid \mu \mid \leq l$. Since the 
spherical harmonics are well known and have been studied thoroughly, we will not comment on them any further. Instead, 
we focus on the radial equation (\ref{radgl}) and its solutions.

\subsection{The Radial Equation}
Before we start constructing the solutions of our radial equation (\ref{radgl}), we will compare it to the one-dimensional 
Schr\"odinger equation studied in \cite{cari,axel1}, and afterwards employ a change of variable to rewrite it in 
dimensionless units.
\subsubsection{Comparison to the one-dimensional case}
In this paragraph we will show that the radial equation (\ref{radgl}) can be cast in a form similar to that of its one-dimensional 
counterpart. To this end, we need to gauge away the $2/r$ term in the coefficient of $R'$ by introduction of a new function 
$u=u(r)$, defined as
\begin{eqnarray}
u&=&r~R,\label{29a}%
\end{eqnarray}
which obeys the following equation
\begin{eqnarray}
(\lambda~r^2+1)~u''+\lambda~r~ u'+
\left[\frac{2~m}{\hbar^2} ~\left(E-V \right)-\frac{(\lambda~r^2+1)~L~(L+1)}{r^2}-\lambda \right] u&=& 0. \label{33}%
\end{eqnarray}
Now define a new energy parameter $\mathcal{E}$ which is shifted from the
energy $E$ by an amount controlled through the parameter $\lambda$:%
\begin{eqnarray}
\mathcal{E}&=&E-\frac{\hbar^{2}\lambda}{2~m},
\label{evalue}%
\end{eqnarray}
and define an effective potential%
\begin{eqnarray}
V_{eff}&=&V+\frac{L~(L+1)~\hbar^2~(\lambda~r^2+1)}{2~m~r^2}. \label{effpot}%
\end{eqnarray}
After substitution of the redefined energy (\ref{evalue}) and potential (\ref{effpot}) into our equation (\ref{33}), 
we obtain its simplified form as
\begin{eqnarray}
(\lambda~r^2+1)~u''+\lambda ~r~u'+\left[  \dfrac
{2~m}{\hbar^{2}} \left(  \mathcal{E}-V_{eff}\right)  \right]  u&=&0.
\label{DE2}%
\end{eqnarray}
More explicitly, using (\ref{pot3d})
\begin{eqnarray}
(\lambda~r^2+1)~u''+\lambda ~r~u'+\left\{  \dfrac
{2~m}{\hbar^{2}} \left[  \mathcal{E}-\frac{1}{2}~g~\frac{r^{2}}{(\lambda~r^2+1)}-
\frac{L~(L+1)~\hbar^2~(\lambda~r^2+1)}{2~m~r^2}\right]  \right\}  u&=&0
\label{DE3}%
\end{eqnarray}
Comparing this to the one-dimensional Schr\"odinger equation of
\cite{cari}, we see that it is the same equation (with $x$ replaced by $r$) for $s$-states, 
where $L=0$. The corresponding solutions for the latter case, and as such for our radial equation in the form 
(\ref{DE3}) for $L=0$, are given by associated Legendre
functions \cite{axel1}. Note also that the energy gets shifted by a constant as displayed in (\ref{evalue}). Let us make 
one more remark on the effective potential (\ref{effpot}). After incorporation of (\ref{pot3d}) and $g=m\alpha^2$ we obtain
\begin{eqnarray}
V_{eff}&=&\frac{1}{2}~m~\alpha^2~\frac{r^{2}}{(\lambda~r^2+1)}+\frac{L~(L+1)~\hbar^2~(\lambda~r^2+1)}{2~m~r^2} \nonumber \\
&=& \frac{1}{2}\left(  \alpha^{2}M~r^{2}+\frac{L~(L+1)~\hbar^2}{M~r^{2}}\right), \label{veffm}
\end{eqnarray}
note that in the last step we used relation (\ref{3}) with $x$ replaced by $r$. We observe that expression (\ref{veffm}) 
exhibits an interesting symmetry, being invariant under the simultaneous
transformations%
\begin{eqnarray}
\alpha^{2}\text{\ }\leftrightarrow~L~(L+1)~\hbar^2 \qquad \qquad
M~r^{2}~\leftrightarrow~\frac{1}{M~r^{2}},\label{35}%
\end{eqnarray}
analysis of which is beyond the scope of this note.

\subsubsection{A dimensionless radial equation}
We will now convert equation (\ref{radgl}) into a dimensionless form, which makes subsequent 
calculations much easier and more transparent. To this end, we define a
dimensionless radial coordinate $y$ and a dimensionless parameter $\Lambda$,
such that $\lambda r^{2}=\Lambda y^{2}$. The coordinates and parameters are
related by%
\begin{eqnarray}
r~=~C~y \qquad \qquad \lambda~=~\frac{\Lambda}{C^{2}}\label{38}%
\end{eqnarray}
where the constant $C$ is to be determined. Using this, we rewrite
Eq.(\ref{radgl}) in the form%
\begin{eqnarray}
(\lambda~y^{2}+1)~R''+\left(  \frac{2}{y}+3~\Lambda ~y\right) R'+\left[\frac{2~m~C^{2}}{\hbar^{2}}\left(  E-V\right) -\frac
{L~(L+1)~(\Lambda~y^{2}+1)}{y^{2}}\right] R&=&0.\label{39}%
\end{eqnarray}
Here it is understood that $R$ is expressed through the new variable $y$, note that 
for reasons of brevity we have not changed the 
function's name. We now look at the term in (\ref{39}) that contains the potential $V$ as a factor. After incorporation of 
(\ref{pot3d}), the latter term reads
\begin{eqnarray}
\frac{2~m~C^{2}}{\hbar^{2}}~V&=& \frac{C^4~g~m~y^2}{\hbar^2~(\lambda~y^2+1)}. \label{40}%
\end{eqnarray}
Once we construct the solutions to our radial equation, it will prove useful to have a redefined constant $g$. 
Adopting the approach taken in \cite{cari}, we set 
\begin{eqnarray}
g &=& m~\alpha^2+\hbar~\alpha~\lambda \nonumber \\[1ex]
&=& m~\alpha^2+\frac{\hbar~\alpha~\Lambda}{C^2}, \nonumber
\end{eqnarray}
where we used our transformation (\ref{38}). This renders the term (\ref{40}) in the following form:
\begin{eqnarray}
\frac{2~m~C^{2}}{\hbar^{2}}~V&=& \frac{C^4~(m^2~\alpha^2+m~\hbar~\alpha~\lambda)~y^2}{\hbar^2~(\lambda~y^2+1)}. 
\label{gx}
\end{eqnarray}
Next, we require $C^{4}m^{2}\alpha^{2}/\hbar^{2}=1$, that is, the free constant $C$ is chosen to be
\begin{eqnarray}
C &=& \sqrt{\frac{\hbar}{m~\alpha}}, \label{c}
\end{eqnarray}
such that expression (\ref{gx}) obtains a simple, dimensionless form:
\begin{eqnarray}
\frac{2~m~C^{2}}{\hbar^{2}}~V&=& \frac{(\Lambda+1)~y^2}{\Lambda~y^2+1}. \label{gxx}
\end{eqnarray}
In the final step we plug (\ref{c}) and (\ref{gxx}) into our rewritten radial equation (\ref{39}), which becomes 
\begin{eqnarray}
(\Lambda~y^2+1)~R''+\left(  \frac{2}{y}+3~\Lambda ~y\right) R'+\left[2~e-\frac{(\Lambda+1)~y^2}{\Lambda~y^2+1}-\frac
{L~(L+1)~(\Lambda~y^{2}+1)}{y^{2}}\right] R&=&0, \label{radfin}
\end{eqnarray}
where the abbreviation $e = E/(\hbar \alpha)$ was employed. We observe that the radial equation in its form 
(\ref{radfin}) is dimensionless.

\subsubsection{Solutions of the radial equation}
In this section we will show that our radial equation (\ref{radfin}) admits a discrete spectrum and a corresponding 
set of orthogonal solutions. In order to simplify our calculations, we choose a system of units where 
$\hbar=1$, $m=1$, and we set the speed of
light $c=1$. In this system of units, the
dimension of distance is the reciprocal of the dimension of mass, both of
which are now dimensionless. Let us now define the following intervals $D_\Lambda \subset \mathbb{R}$:
\begin{eqnarray}
D_\Lambda &=& \left\{
\begin{array}{lll}
\left(0,\sqrt{\frac{1}{|\Lambda|}} \right) & \mbox{if} & \Lambda<0 \\[2ex]
(0,\infty) & \mbox{if} & \Lambda >0 
\end{array}
\right\}. \label{dlambda}
\end{eqnarray}
The case $\Lambda=0$ corresponds to the well-known harmonic oscillator model, as can be inferred from inspection of 
(\ref{radfin}), for details the reader may refer to \cite{messiah}. Now, 
the boundary-value problem for our radial equation (\ref{radfin}), defined on $D_\Lambda$, takes the following form
\begin{eqnarray}
(\Lambda~y^2+1)~R''+\left(\frac{2}{y}+3~\Lambda~y \right) R'+
\left[2~e-L(L+1)~\Lambda-1+\frac{1-y^2}{\Lambda~y^2+1}-\frac{L(L+1)}{y^2} \right] R
&=&0, \nonumber \\ \label{eq} 
\end{eqnarray}
equipped with $\Lambda$-dependent boundary conditions:
\begin{eqnarray}
R(0) < \infty \qquad \mbox{and} \qquad
\left\{
\begin{array}{lll}
R\left(\sqrt{\frac{1}{|\Lambda|}}\right) = 0 &\mbox{if}& \Lambda<0 \\[2ex]
\lim\limits_{y \rightarrow \infty} R(y) = 0 &\mbox{if}& \Lambda>0 
\end{array}
\right\}. \label{bc}
\end{eqnarray}
Note that we have expanded the coefficient of $R$ in (\ref{radfin}). Relations (\ref{eq}), (\ref{bc}) form a spectral problem with $e$ being the spectral parameter. In order to be 
physically meaningful, solutions of equation (\ref{eq}) must be located in the following weighted Hilbert space
\begin{eqnarray}
L^2_\mu(D_\Lambda) &=& \left\{f:D_\Lambda \rightarrow \mathbb{C}\mid \int\limits_{D_\Lambda} \vert f \vert^2~\mu~dy<\infty
\right\}. \label{l2}
\end{eqnarray} 
The weight function $\mu$ that appears in the latter definition is given by
\begin{eqnarray}
\mu = \frac{y^2}{\sqrt{\Lambda~y^2 +1}}. \label{mu}
\end{eqnarray}
Note that the numerator of this weight function is due to the measure in spherical coordinates. Let us finally remark that 
the internal product of the space (\ref{l2}), that induces its norm, is defined as
\begin{eqnarray}
\left(f,g\right)_{L^2_\mu} &=& \int\limits_{D_\Lambda} f^\ast g~\mu~dy, \nonumber
\end{eqnarray}
where the asterisk stands for complex conjugation. Equation 
(\ref{eq}) can be solved exactly, its general solution has the form
\begin{eqnarray}
R &=& c_1~F+c_2~G, \label{sol}
\end{eqnarray}
where the two functions $F$ and $G$ are given by the following expressions
\begin{eqnarray}
F &=& y^{L} \left(\Lambda~y^2+1 \right)^{-\frac{1}{2 \Lambda}} 
{}_2F_1\left(\alpha-\beta,\alpha+\beta,L+\frac{3}{2},-\Lambda~y^2 \right) \label{r1} \\
G &=& y^{-L-1} \left(\Lambda~y^2+1 \right)^{\frac{1}{2 \Lambda}} 
{}_2F_1\left(\alpha-L-\frac{1}{2}+\beta,\alpha-L-\frac{1}{2}-\beta,-L+\frac{1}{2},-\Lambda~y^2 \right).
\label{r2}
\end{eqnarray}
Here, ${}_2F_1$ stands for the hypergeometric function, the arguments of which contain the following 
parameters
\begin{eqnarray}
\alpha &=& \frac{L}{2}+\frac{1}{2}-\frac{1}{2~\Lambda} 
\label{alpha} \\[1ex]
\beta &=& \frac{\sqrt{\Lambda+1-2~e~\Lambda+\Lambda^2 (L^2+L+1)}}{2~\Lambda}. \label{beta}
\end{eqnarray}
Since $L$ is a nonnegative integer and since the hypergeometric function behaves like a constant around the origin, 
the solution (\ref{r2}) is in general singular there and as such does not satisfy our first boundary condition in (\ref{bc}). 
Consequently, we must discard the solution (\ref{r2}) by setting $c_2=0$ in (\ref{sol}). Focusing on the remaining 
solution (\ref{r1}), we first recall that the hypergeometric function behaves asymptotically like an exponential function. Since 
this would lead to an unbounded solution and to a violation of our second boundary condition in (\ref{bc}), we will convert the hypergeometric function into 
a polynomial by requesting its first argument to equal a negative integer or zero. More precisely, we impose the condition 
$\alpha-\beta = -n$, where $n \in \mathbb{N} \cup \{0\}$. Solving the latter condition for $e$ leads to a discrete set $(e_n) \subset 
\mathbb{R}$ of energies:
\begin{eqnarray}
e_n &=& -2~\Lambda~n^2-2~L~\Lambda~n-2~\Lambda~n-\frac{L~\Lambda}{2}+2~n+L+\frac{3}{2}. 
\label{ene}
\end{eqnarray}
We will now substitute this relation into our parameters (\ref{alpha}), (\ref{beta}) and plug them into the solution 
(\ref{r1}). For the sake of simplicity, we will identify the functions $F$ and $R$, since they differ only by an irrelevant 
constant. Thus, we obtain the following set of solutions $(R_n)$, $n \in \mathbb{N} \cup \{0\}$, 
\begin{eqnarray}
R_n &=& y^{L} \left(\Lambda~y^2+1 \right)^{-\frac{1}{2 \Lambda}} 
{}_2F_1\left(-n,n+L+1-\frac{1}{\Lambda},L+\frac{3}{2},-\Lambda~y^2 \right). \label{r1s}
\end{eqnarray}
Due to its first argument being a nonpositive integer, the hypergeometric function degenerates to a 
polynomial. In particular, the solution (\ref{r1s}) can be rewritten as the set of functions $(R_n)$, 
$n \in \mathbb{N} \cup \{0\}$, given by
\begin{eqnarray}
R_n &=&  y^{L} \left(\Lambda~y^2+1 \right)^{-\frac{1}{2 \Lambda}}
P_n^{\left(L+\frac{1}{2},-\frac{1}{\Lambda}-\frac{1}{2} \right)}\left(1+2~\Lambda~y^2\right), \label{solfin}
\end{eqnarray}
note that $P$ stands for a Jacobi polynomial and that irrelevant factors have been discarded. We will now show that 
the solution set (\ref{r1s}) is orthogonal and that its elements are normalizable. To this end, we distinguish the two cases of negative and 
positive $\Lambda$.

\paragraph{First case: \boldmath{$\Lambda<0$.}} It will prove convenient to work with the representation (\ref{solfin}) of 
our solutions, which for negative $\Lambda$ can be rewritten as follows
\begin{eqnarray}
R_n &=&  y^{L} \left(-|\Lambda|~y^2+1 \right)^{\frac{1}{2 |\Lambda|}}
P_n^{\left(L+\frac{1}{2},\frac{1}{|\Lambda|}-\frac{1}{2} \right)}\left(1-2~|\Lambda|~y^2\right). \label{solfinx}
\end{eqnarray}
In the first step we will show that these functions fulfill our boundary conditions (\ref{bc}) and are normalizable. 
To this end, we observe that the solutions (\ref{solfinx}) are continuous and bounded on $D_\Lambda$, such that 
the first boundary condition verifies. Next, it is straightforward to see that the functions (\ref{solfinx}) always 
admit a zero at $r=\sqrt{1/|\Lambda|}$, because the power of the second factor on the right-hand side is always positive. 
In conclusion, all functions (\ref{solfinx}) comply with our boundary conditions (\ref{bc}). In order to investigate 
normalizability, we notice that the corresponding norm $\Vert \cdot \Vert_{L_\mu^2}$ is obtained by integrating 
the following expression over $D_\Lambda$:
\begin{eqnarray}
|R_n|^2 ~\mu &=& |R_n|^2 \frac{y^2}{\sqrt{-|\Lambda|~y^2 +1}}, \label{inte}
\end{eqnarray}
where the weight function $\mu$ was defined in (\ref{mu}) and has been adjusted to negative $\Lambda$. Since 
expression (\ref{inte}) is continuous on $D_\Lambda$, we must verify boundedness at the endpoints of the latter interval. 
Inspection of (\ref{solfinx}) and (\ref{inte}) shows that $|R_n|^2\mu$ behaves at $y=0$ like $y^{2L+2}$, and as such stays 
bounded there. At the remaining endpoint $y=\sqrt{1/|\Lambda|}$ of our domain we find the asymptotics 
$(-|\Lambda|~y^2+1 )^{\frac{1}{|\Lambda|}-\frac{1}{2}}$. We see that although the integrand (\ref{inte}) becomes 
unbounded for $\Lambda<-2$, it remains integrable, because its exponent stays above -1. Next, in order to show that the solution 
family $(R_n)$ is orthogonal with respect to the inner product in 
$L^2_\mu(D_\Lambda)$, we make use of the following, well-known identity \cite{abram}
\begin{eqnarray}
\int\limits_{-1}^1 (1-x)^a ~(1+x)^b~ P_m^{(a,b)}(x)~P_n^{(a,b)}(x)~dx &=& c~\delta_{mn},
\label{ortho}
\end{eqnarray}
valid for $a,b \in \mathbb{R}$, $a,b > -1$, and $m,n \in \mathbb{N} \cup \{0\}$. Furthermore, 
$\delta$ denotes the Kronecker symbol and $c \in \mathbb{C}$ is a constant, the explicit form of which 
is not of importance here. Let us apply the following settings to relation (\ref{ortho}):
\begin{eqnarray}
a~=~L+\frac{1}{2} \qquad \qquad b~=~\frac{1}{|\Lambda|}-\frac{1}{2} \qquad \qquad 
x ~=~ 1-2~ |\Lambda|~y^2, \label{abl}
\end{eqnarray}
It is immediate to see that $a,b>-1$, such that (\ref{ortho}) is applicable. Absorbing 
all constant factors into a parameter $k \in \mathbb{C}$, the left-hand side of (\ref{ortho}) 
obtains the following form
\begin{eqnarray}
& & \hspace{-0.5cm}\int\limits_{-1}^1 (1-x)^a ~(1+x)^b~ P_m^{(a,b)}(x)~P_n^{(a,b)}(x)~dx ~=~ \nonumber \\
& & =~
k~\int\limits_0^{\sqrt{\frac{1}{|\Lambda|}}} y^{2L+2} 
\left(-|\Lambda|~y^2 +1\right)^{\frac{1}{|\Lambda|}-\frac{1}{2}} 
P_m^{\left(L+\frac{1}{2},\frac{1}{|\Lambda|}-\frac{1}{2} \right)}\left(1-2~ |\Lambda|~y^2\right)
P_n^{\left(L+\frac{1}{2},\frac{1}{|\Lambda|}-\frac{1}{2} \right)}\left(1-2~ |\Lambda|~y^2\right) dy \nonumber \\[1ex]
& & =~
k~\int\limits_{D_\Lambda} R_m^2~R_n^2~
\frac{y^2}{\sqrt{-|\Lambda|~y^2 +1}}~dy \nonumber \\[1ex]
& & =~k~\left(R_m,R_n \right)_{L^2_\mu}. \nonumber
\end{eqnarray}
Combination of this result with (\ref{ortho}) gives the identity
\begin{eqnarray}
\left(R_m,R_n \right)_{L^2_\mu} &=& C~\delta_{mn}, \nonumber
\end{eqnarray}
where $C=c/k \in \mathbb{C}$. In summary, for $\Lambda <0$, 
the solution of our boundary-value problem (\ref{eq}), (\ref{bc}) is given by 
the infinite number of functions $R_n \in L_\mu^2(D_\Lambda)$, $n \in \mathbb{N} \cup \{0\}$, as displayed in (\ref{solfinx}). In particular, 
the set $(R_n)$ is orthogonal with respect to the inner product of $L_\mu^2(D_\Lambda)$. 
The corresponding spectral values $(e_n)$, $n \in \mathbb{N} 
\cup[ \{0\}$ can be found in (\ref{ene}):
\begin{eqnarray}
e_n &=& 2~|\Lambda|~n^2+2~L~|\Lambda|~n+2~|\Lambda|~n+\frac{L~|\Lambda|}{2}+2~n+L+\frac{3}{2}, 
\label{ene2}
\end{eqnarray}
where the substitution $\Lambda = -|\Lambda|$ was made.

\paragraph{Second case: \boldmath{$\Lambda>0$.}} As in the previous case, we start out by verifying that our solutions 
(\ref{solfin}) satisfy the boundary conditions (\ref{bc}) for the present case of positive $\Lambda$. Inspection of the 
explicit form (\ref{solfin}) shows that the 
functions $R_n$, $n \in \mathbb{N} \cup \{0\}$, behave like $y^L$ close to $y=0$, such that our first boundary 
condition in (\ref{bc}) is fulfilled. In a similar manner we can find our solutions' asymptotics at infinity, which 
turns out to be $r^{L-1/\Lambda+2n}$. If this expression tends to zero, then our second boundary condition in 
(\ref{bc}) is fulfilled, such that we are left with a constraint on our solution index $n$, that is, 
\begin{eqnarray}
n &<& \frac{1}{2~\Lambda}-\frac{L}{2}. \label{ncond1}
\end{eqnarray} 
This constraint already indicates that the number of physically meaningful solutions cannot be infinitely large, as in the 
previous case of negative $\Lambda$. Before we comment on this matter in more detail, let us verify normalizability 
of our solutions, assuming the above condition on $n$ to be fulfilled. The norm $\Vert \cdot \Vert_{L^2_\mu}$ 
of our solutions is obtained by integrating the expression
\begin{eqnarray}
|R_n|^2 ~\mu &=& |R_n|^2 \frac{y^2}{\sqrt{\Lambda~y^2 +1}}, \label{inte1}
\end{eqnarray}
over $D_\Lambda$, which in the present case is the positive real axis. Inspection of our solutions (\ref{solfin}) shows that 
(\ref{inte1}) is continuous on $D_\Lambda$, such that only the asymptotics of (\ref{inte1}) at infinity can affect 
normalizability of the functions (\ref{solfin}). It is straightforward to verify that (\ref{inte1}) 
behaves like $y^{2L+1-2/\Lambda+4n}$ at infinity, which results in the condition
\begin{eqnarray}
n &<&  \frac{1}{2~\Lambda}-\frac{1}{2}-\frac{L}{2}. \label{ncond2}
\end{eqnarray}
We see that this condition on $n$ is stronger than its previous counterpart (\ref{ncond1}), such that we can 
work solely with (\ref{ncond2}). If the latter condition is satisfied, then the corresponding function (\ref{inte1}) can be 
integrated over $D_\Lambda$. In other words, $R_n \in L^2_\mu(D_\Lambda)$ for all $n$ that 
(\ref{ncond2}) applies to. Hence, in contrast to the previous case of negative $\Lambda$, the number of physically 
acceptable solutions to our boundary-value problem (\ref{eq}), (\ref{bc}) is limited and depends on $\Lambda$. 
In particular, if the relation
\begin{eqnarray}
\Lambda &\geq& \frac{1}{1+L}, \nonumber
\end{eqnarray}
holds, then there is no physical solution at all. Now that we have completed our analysis on normalizability, it 
remains to show that the functions (\ref{solfin}) form an orthogonal set with respect to the inner product on 
$L_\mu^2(D_\Lambda)$. To this end, we will work with the hypergeometric representation (\ref{r1s}) of our 
solutions. Taking into account that the latter functions are real-valued, we obtain 
for $n,m \in \mathbb{N} \cup \{0\}$ and $m \neq n$, the internal product of two different solutions as
\begin{eqnarray}
\left(R_m,R_n \right)_{L^2_\mu} &=& \int\limits_{D_\Lambda} R_m^\ast ~R_n~\mu~dy  \nonumber \\
&=& \int\limits_0^\infty 
y^{2L+2} \left(\Lambda~y^2+1 \right)^{-\frac{1}{\Lambda}-\frac{1}{2}} 
{}_2F_1\left(-m,m+L+1-\frac{1}{\Lambda},L+\frac{3}{2},-\Lambda~y^2 \right) \times \nonumber \\
&\times& 
{}_2F_1\left(-n,n+L+1-\frac{1}{\Lambda},L+\frac{3}{2},-\Lambda~y^2 \right) dy. \label{o1}
\end{eqnarray}
Note that our weight function (\ref{mu}) has been absorbed into the first two factors of the integral. In the next step we 
will employ the following change of variable:
\begin{eqnarray}
y &=& \sqrt{\frac{1-x}{\Lambda~(1+x)}}. \label{change}
\end{eqnarray}
It is straightforward to see that this function maps $x \in (-1,1)$ onto $y \in (0,\infty)$. The differential of 
(\ref{change}) is given by
\begin{eqnarray}
dy &=& \frac{1}{x^2-1}~\sqrt{\frac{1-x}{\Lambda~(1+x)}}~dx, \nonumber
\end{eqnarray}
such that after substitution of the latter relation and (\ref{change}), our inner product (\ref{o1}) takes the following form:
\begin{eqnarray}
\left(R_m,R_n \right)_{L^2_\mu} &=& a_1~\int\limits_{-1}^1 (1-x)^{L+\frac{1}{2}}~ (1+x)^{-2-L+\frac{1}{\Lambda}}~
{}_2F_1\left(-m,m+L+1-\frac{1}{\Lambda},L+\frac{3}{2},\frac{x-1}{x+1}\right) \times \nonumber \\
&\times& 
{}_2F_1\left(-n,n+L+1-\frac{1}{\Lambda},L+\frac{3}{2}, \frac{x-1}{x+1}\right) dx. \label{o2}
\end{eqnarray}
Note that all irrelevant constant factors have been absorbed into $a_1 \in \mathbb{R}$. In the next step we convert the 
hypergeometric functions into Jacobi polynomials, using the following identity \cite{ryzhik}
\begin{eqnarray}
{}_2F_1\left(-n,-n-b,a+1,\frac{x-1}{x-1} \right) &=& \left(\frac{2}{x+1} \right)^{n} \frac{\Gamma(a+1)~n!}{\Gamma(n+a+1)}~
P_n^{(a,b)}(x), \label{jac}
\end{eqnarray}
valid for $n \in \mathbb{N} \cup \{0\}$ and $a,b,x \in \mathbb{R}$, provided the hypergeometric function exists. In order to 
use relation (\ref{jac}) in the present case, we must choose $a=L+1/2$ and $b=-2n-L-1+1/\Lambda$ resp. 
$b=-2m-L-1+1/\Lambda$, as inspection of (\ref{o2}) shows. The 
inner product then renders in the form
\begin{eqnarray}
\left(R_m,R_n \right)_{L^2_\mu} &=& a_2~\int\limits_{-1}^1 (1-x)^{L+\frac{1}{2}}~ (1+x)^{-2-L+\frac{1}{\Lambda}-m-n}~
P_m^{\left(L+\frac{1}{2},-1-L-2m+\frac{1}{\Lambda} \right)}(x) \times \nonumber \\
&\times& P_n^{\left(L+\frac{1}{2},-1-L-2n+\frac{1}{\Lambda} \right)}(x)~dx, \label{o3}
\end{eqnarray}
where constant factors from (\ref{jac}) have been put into $a_2 \in \mathbb{R}$. We observe that the internal product 
(\ref{o3}) is invariant under exchanging $m$ and $n$. Without restriction, we can therefore assume that $m>n$ and 
replace the Jacobi polynomial of order $m$ by its representation through Rodrigues' relation \cite{abram}:
\begin{eqnarray}
P_m^{(a,b)} &=& \frac{(-1)^m}{2^m ~m!}~(1-x)^{-a}~(1+x)^{-b}~\frac{d^m}{dx^m} 
\left[ (1-x)^{a+m}~(1+x)^{b+m} \right]. \label{rod}
\end{eqnarray}
where we must use the parameter settings $a=L+1/2$ and $b=-2m-L-1+1/\Lambda$:
\begin{eqnarray}
\left(R_m,R_n \right)_{L^2_\mu} &=& a_3~\int\limits_{-1}^1 (1+x)^{-1+m-n}~\frac{d^m}{dx^m} 
\left[ (1-x)^{L+\frac{1}{2}+m}~(1+x)^{-m-L-1+\frac{1}{\Lambda}} \right]
\times \nonumber \\
&\times& P_n^{\left(L+\frac{1}{2},-1-L-2n+\frac{1}{\Lambda} \right)}(x)~dx, \label{o4}
\end{eqnarray}
note that constant factors from (\ref{rod}) have been placed into $a_3 \in \mathbb{R}$. In the next step, we 
perform an $n$-fold integration by parts, which yields
\begin{eqnarray}
\left(R_m,R_n \right)_{L^2_\mu} &=& a_3~\int\limits_{-1}^1 
\frac{d^m}{dx^m} \left[ (1+x)^{-1+m-n}~P_n^{\left(L+\frac{1}{2},-1-L-2n+\frac{1}{\Lambda} \right)}(x) \right] \times \nonumber \\
&\times&
(1-x)^{L+\frac{1}{2}+m}~(1+x)^{-m-L-1+\frac{1}{\Lambda}}~dx. \label{o5}
\end{eqnarray}
Let us point out that the intermediate terms of each integration by parts vanish due to nonnegative exponents. In particular, 
the last exponent under the derivative in (\ref{o4}) is positive, because
\begin{eqnarray}
-m-L-1+\frac{1}{\Lambda} &>& -2~m-L-\frac{1}{2}+\frac{1}{\Lambda} ~~>~~ 0, \nonumber
\end{eqnarray}
due to our assumption (\ref{ncond2}). Our final step consists in observing that in (\ref{o5}) the $m$-fold 
derivative yields zero, because it is applied to a polynomial of degree $-1+m-n+n=m-1$. Therefore, we conclude
\begin{eqnarray}
\left(R_m,R_n \right)_{L^2_\mu} &=& a_3~\delta_{mn}, \nonumber
\end{eqnarray}
which establishes orthogonality. In summary, for $\Lambda \in (0,\infty)$, 
the solution of our boundary-value problem (\ref{eq}), (\ref{bc}) is given by 
the functions $R_n \in L_\mu^2(D_\Lambda)$, $n \in \mathbb{N} \cup \{0\}$, as displayed in (\ref{solfinx}). The number of 
functions is limited and depends on $\Lambda$ through relation (\ref{ncond2}). The set $(R_n)$ is orthogonal with 
respect to the inner product of $L_\mu^2(D_\Lambda)$ and the corresponding spectral values $(e_n)$, $n \in \mathbb{N} 
\cup[ \{0\}$, are given by (\ref{ene}).

\subsubsection{Convergence to the harmonic oscillator}
We will now briefly verify that the solutions and spectral values found in the previous section reduce correctly to their 
well-known counterparts if our parameter $\Lambda$ goes to zero. Since the angular part of our model does not involve 
$\Lambda$, it is sufficient to study the radial boundary-value problem (\ref{eq}), (\ref{bc}) and its domain $D_\Lambda$, 
as given in (\ref{dlambda}). Starting out with the latter domain, we observe that its two distinct definitions for 
negative and positive $\Lambda$ become the same in the limiting case, that is, 
$\lim_{\Lambda \rightarrow 0} D_\Lambda = (0,\infty)$. Applying the same limit to the radial equation (\ref{eq}), our 
boundary-value problem reads
\begin{eqnarray}
R''+\frac{2}{y}~R'+
\left[2~e-y^2-\frac{L(L+1)}{y^2} \right] R
&=&0, ~~~y \in (0,\infty) \label{eq2} \\[1ex]
R(0) ~<~ \infty,~~
\lim\limits_{y \rightarrow \infty} R(y) &=& 0. \label{bc2}
\end{eqnarray}
This coincides with the well-known radial equation for the three-dimensional harmonic oscillator \cite{messiah}. After 
having determined convergence of the boundary-value problem, let us now investigate its solutions. As far as the 
corresponding spectral values are concerned, their explicit forms (\ref{ene}) and (\ref{ene2}) tend to the same limit 
if $\Lambda$ goes to zero:
\begin{eqnarray}
e_n &=& 2~n+L+\frac{3}{2}, \label{eneosc}
\end{eqnarray}
which gives precisely the energy spectrum of the harmonic oscillator in three dimensions \cite{messiah}. In the final 
step we must apply the limit $\Lambda \rightarrow 0$ to our solutions, where we will take the representation (\ref{r1s}). 
Using standard identities \cite{abram}, we find
\begin{eqnarray}
\lim\limits_{\Lambda \rightarrow 0} ~y^{L} \left(\Lambda~y^2+1 \right)^{-\frac{1}{2 \Lambda}} 
{}_2F_1\left(-n,n+L+1-\frac{1}{\Lambda},L+\frac{3}{2},-\Lambda~y^2 \right) ~= \nonumber \\[1ex]
& & \hspace{-5cm}=~ y^L \exp\left(-\frac{1}{2}~ y^2 \right) 
{}_1F_1\left(-n,L+\frac{3}{2},y^2 \right) \nonumber \\[1ex]
& &\hspace{-5cm}=~y^L \exp\left(-\frac{1}{2}~ y^2 \right)~L_n^{L+\frac{1}{2}}(y^2), \label{solosc}
\end{eqnarray}
where some irrelevant factors have been omitted. Note that ${}_1F_1$ stands for the confluent hypergeometric function 
and $L$ represents a generalized Laguerre polynomial. 
The functions (\ref{solosc}) are the well-known solutions of the boundary-value problem (\ref{eq2}), (\ref{bc2}) that lie in 
$L^2(0,\infty)$ and belong to the discrete spectral values given by (\ref{eneosc}).

\section{Concluding remarks}
In this note we have studied the quantum model of an isotropic nonlinear oscillator in three spatial dimensions. 
Via separation of variables in the underlying Schr\"odinger equation, we obtained the entire discrete spectrum of the 
problem, together with a corresponding set of orthogonal solutions. Connections to the one-dimensional situation and 
the conventional harmonic oscillator system were established. A natural continuation of this work could concern the 
behaviour of our model in a setting of spherical symmetry and arbitrary dimension (hyperspherical symmetry). 
In such a case, properties of our system like the energy spectrum and the number of physically acceptable solutions become 
functions of the dimension.

\end{document}